\documentclass[12pt,english]{article}
\pdfoutput=1
\usepackage{lmodern}

\usepackage[T1]{fontenc}
\usepackage[latin9]{inputenc}
\usepackage[numbers,sort&compress]{natbib}
\usepackage{geometry}
\usepackage{color}
\usepackage{babel}
\usepackage{mathtools}
\usepackage{amsmath}
\usepackage{amssymb}
\usepackage{graphicx}
\usepackage{esint}
\usepackage[unicode=true,pdfusetitle,
 bookmarks=true,bookmarksnumbered=false,bookmarksopen=false,
 breaklinks=false,pdfborder={0 0 1},backref=false,colorlinks=true]
 {hyperref}
\hypersetup{
 citecolor=black,linkcolor=black,urlcolor=black}
\makeatletter

 \@ifundefined{textcolor}{}
 {
   \definecolor{BLACK}{gray}{0}
   \definecolor{WHITE}{gray}{1}
   \definecolor{RED}{rgb}{1,0,0}
   \definecolor{GREEN}{rgb}{0,1,0}
   \definecolor{BLUE}{rgb}{0,0,1}
   \definecolor{CYAN}{cmyk}{1,0,0,0}
   \definecolor{MAGENTA}{cmyk}{0,1,0,0}
   \definecolor{YELLOW}{cmyk}{0,0,1,0}
 }
\usepackage{babel}
\makeatletter
\def\simgt{\mathrel{\lower2.5pt\vbox{\lineskip=0pt\baselineskip=0pt
           \hbox{$>$}\hbox{$\sim$}}}}
\def\simlt{\mathrel{\lower2.5pt\vbox{\lineskip=0pt\baselineskip=0pt
           \hbox{$<$}\hbox{$\sim$}}}}
\makeatother

\newcommand{\bea}{\begin{eqnarray}}
\newcommand{\eea}{\end{eqnarray}}
\newcommand{\eq}[1]{\begin{align}#1\end{align}}

\newcommand{\Fig}[1]{Fig.~\ref{#1}}
\newcommand{\Eq}[1]{Eq.~\eqref{#1}}

\newcommand{\Sec}[1]{Sec.~\ref{#1}}

\newcommand{\ee}[2]{e_{#1}e_{#2}}
\newcommand{\pp}[2]{p_{#1}p_{#2}}
\newcommand{\pe}[2]{p_{#1}e_{#2}}

\begin{document}
	\interfootnotelinepenalty=10000
	\baselineskip=18pt
	\hfill CALT-TH-2017-22
	\hfill
	
	\vspace{2cm}
	\thispagestyle{empty}
	\begin{center}
		{\LARGE 
			Unifying Relations for Scattering Amplitudes
		}\\
		\bigskip\vspace{1.cm}{
			{\large Clifford Cheung,${}^a$ Chia-Hsien Shen,${}^a$ and Congkao Wen${}^{a,b}$}
		} \\[7mm]
		{\it  ${}^a$Walter Burke Institute for Theoretical Physics, \\[-1mm]
			California Institute of Technology, Pasadena, CA 91125
			\\
			${}^b$Mani L. Bhaumik Institute for Theoretical Physics, \\[-1mm]
			 Department of Physics and Astronomy, UCLA, Los Angeles, CA 90095
			}\let\thefootnote\relax\footnote{e-mail: \url{clifford.cheung@caltech.edu}, \url{chshen@caltech.edu}, \url{cwen@caltech.edu }} \\
	\end{center}
	\bigskip
	\centerline{\large\bf Abstract}

\begin{quote} \small
	We derive new amplitudes relations revealing a hidden unity among a wide-ranging variety of  theories in arbitrary spacetime dimensions.  Our results rely on a set of Lorentz invariant differential operators which transmute physical tree-level scattering amplitudes into new ones.   By transmuting the amplitudes of gravity coupled to a dilaton and two-form, we generate all the amplitudes of Einstein-Yang-Mills theory, Dirac-Born-Infield theory,
 special Galileon, nonlinear sigma model, and biadjoint scalar theory. Transmutation also relates amplitudes in string theory and its variants.
	As a corollary, celebrated aspects of gluon and graviton scattering like color-kinematics duality, the KLT relations, and the CHY construction are inherited traits of the transmuted amplitudes.  Transmutation recasts the Adler zero as a trivial consequence of the Weinberg soft theorem and implies new subleading soft theorems for certain scalar theories.
\end{quote}

\setcounter{footnote}{0}

\newpage
\tableofcontents
	
\newpage

\section{Introduction and Summary}

The modern S-matrix program has exposed marvelous structures long hidden in plain sight within gauge theory and gravity. The fact that these theories are endowed with exceptional properties is perhaps unsurprising given that their form is uniquely dictated by locality and gauge invariance.  Nevertheless, in recent years it has become abundantly clear that many of these structures are actually commonplace.  For example, the Kawai-Lewellen-Tye (KLT) relations \cite{Kawai:1985xq}, Bern-Carrasco-Johansson (BCJ) color-kinematics duality \cite{Bern:2008qj}, and Cachazo-He-Yuan (CHY) construction \cite{Cachazo:2013hca,Cachazo:2013iea, Cachazo:2014xea} all apply across a tremendous range of theories, including Einstein-Yang-Mills (EYM) theory, Dirac-Born-Infeld (DBI) theory, the special Galileon (SG), the nonlinear sigma model (NLSM), and the biadjoint scalar (BS) theory.   
A natural question then arises: why {\it these} theories?   Imbued with symmetries of disparate origin and character, these theories do not obviously conform to any cohesive organizing principle that would place them on equal footing in the eyes of the S-matrix.

In this paper, we offer an explanation for the peculiar universality of these structures.  Our results follow from a set of simple unifying relations which ``transmute'' the tree-level scattering amplitudes of certain theories into those of others. By transmuting the S-matrix of the mother of all theories---gravity coupled to a dilaton and two-form, {\it i.e.},~``extended gravity''---we beget the S-matrices of all the theories previously mentioned, thus revealing their many shared traits as congenital.

\begin{figure*}[t]
	\begin{center}
		\includegraphics[width=0.75\textwidth]{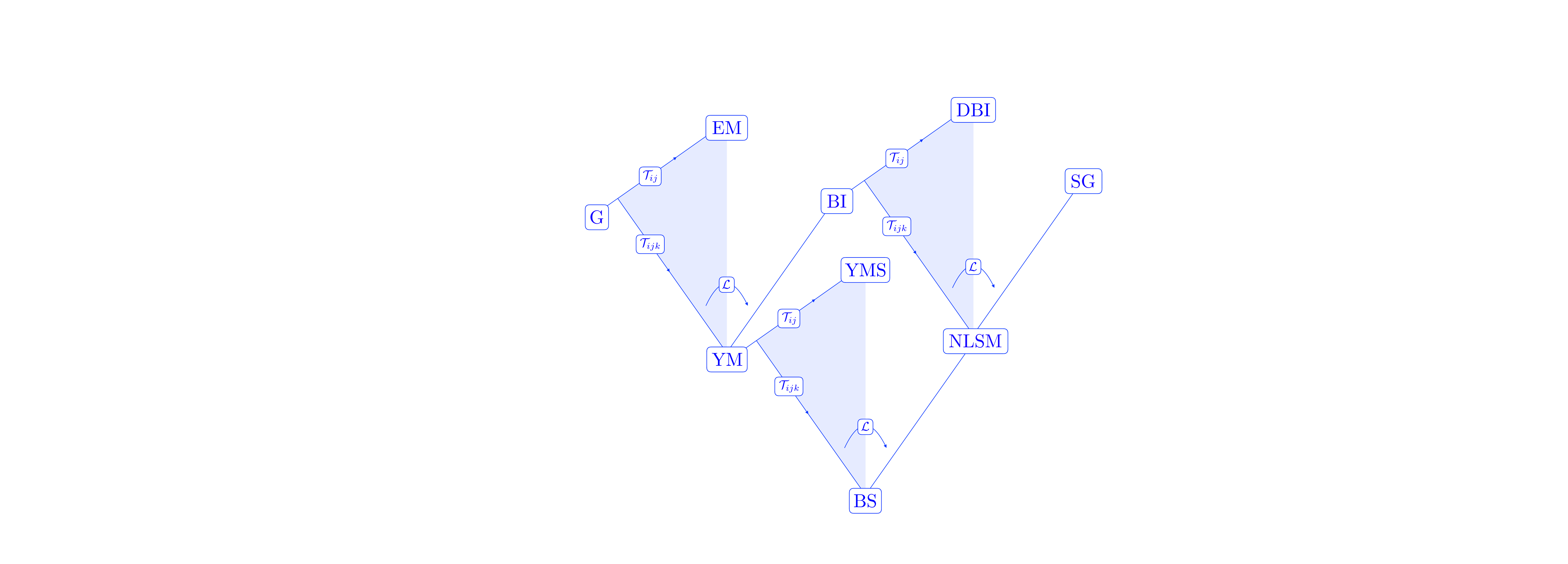}
	\end{center}
	\vspace{-5mm}
	\caption{Diagram depicting the unified web of  theories.   The corners represent extended gravity (G), Einstein-Maxwell (EM) theory, Yang-Mills (YM) theory, Born-Infeld (BI) theory, Dirac-Born-Infeld (DBI) scalar theory, nonlinear sigma model (NLSM), special Galileon (SG), Yang-Mills scalar (YMS) theory, and biadjoint (BS) theory.  The arrows correspond to the transmutation operators, ${\cal T}_{ij}$, ${\cal T}_{ijk}$, and ${\cal L}$.  The shaded regions and edges correspond to hybrid theories.}
	\label{fig:unified_web}
\end{figure*}

Transmutation is by design independent of the particular form in which an amplitude happens to be represented. To achieve this, we study amplitudes as formal functions of Lorentz invariant products of the external on-shell kinematic data: $e_i e_j$, $p_i  e_j$, and $p_i  p_j$ where $e_i$ and $p_i$ denotes polarization and momentum of particle $i$, respectively, and $i\neq j$.\footnote{So as not to complicate our notation we omit the usual ``$\cdot$'' denoting Lorentz-invariant contractions.}
\footnote{In the extended gravity, all the degrees of freedom can be incorporated into a single polarization tensor $e_{\mu\nu}$. It can written as a product of two copies of YM polarizations, $e_{\mu\nu}=e_{\mu}\tilde{e}_{\nu}$. The usual pure graviton corresponds to the symmetric and traceless component of $e_{\mu\nu}$. Since the transmutation only acts on one of the copies, we focus on the un-tilded copy without loss of generality.}  In this language, it is straightforward to construct a basis of gauge invariant operators which preserve on-shell kinematics.  Perhaps unsurprisingly, these operators automatically transmute gauge invariant scattering amplitudes into new ones.  Our roster of transmutation operators is summarized as follows:
\begin{itemize}
	\item The trace operator ${\cal T}_{ij}=\partial_{e_i e_j}$ reduces the spin of particles $i$ and $j$ by one unit and places them within a new color trace structure. This operator transmutes gravitons into photons,  gluons into biadjoint scalars, and BI photons into DBI scalars.

	\item The insertion operator ${\cal T}_{ijk} = \partial_{p_i e_j}- \partial_{p_ke_j}$ reduces the spin of particle $j$ by one unit and inserts it between particles $i$ and $k$ within a color trace structure. This operator transmutes gravitons into gluons, gluons into biadjoint scalars, and BI photons into pions, at the level of color-ordered amplitudes of the resulting theories, which we will properly define in \Sec{sec:web}.
	
	\item The longitudinal operator ${\cal L}_i = \sum_j p_i p_j \partial_{p_j e_i}$ reduces the spin of particle $i$ by one unit while converting it to a longitudinal mode.  Necessarily applied to all particles at once, this operator transmutes gravitons into BI photons, gluons into pions, and BI photons into SG scalars.
\end{itemize}
These transmutation operators form the building blocks of the unified web of theories depicted in \Fig{fig:unified_web}.   Theory space is partitioned into three triangular regions corresponding to the nearest of kin to extended gravity, BI, and YM.
The corners of the web denote all of the theories previously discussed, while the edge and bulk regions denote  hybrid theories.  Remarkably, these triangular regions are not independent---due to KLT \cite{Kawai:1985xq} and BCJ \cite{Bern:2008qj}, extended gravity can be recast as the double copy of YM, and BI as the product of YM and the NLSM.  Hence, one can actually construct the entire web purely from the middle triangle corresponding to YM and its descendants.


Armed with the notion of transmutation, it is then straightforward to import familiar structures in gauge theory and gravity into their descendants.  For example, transmutation trivially explains the near-universal applicability of the KLT, BCJ, and CHY constructions.  Likewise, the Weinberg soft theorems transmute into the Adler zero conditions on pions and Galileons, as first observed in \cite{Cheung:2016prv}.  At subleading order, we can also derive new soft theorems in the spirit of \cite{Cachazo:2016njl}.

The remainder of this paper is organized as follows.   In Sec.~\ref{sec:setup}, we systematically construct a basis of transmutation operators. 
We define the notion of color-ordering which is crucial for insertion and discuss the web of theories related by transmutation in Sec.~\ref{sec:web}. Examples are illustrated in Sec.~\ref{sec:example}.
A proof by induction of our claims is presented in Sec.~\ref{sec:proof}, followed by a discussion of the infrared structure in Sec.~\ref{sec:soft}.  Finally, we conclude with a discussion of  outlook and future directions in Sec.~\ref{sec:outlook}.

\section{Transmutation}
\label{sec:setup}

\subsection{Physical Constraints}

In an abstract sense, the purpose of a transmutation operator ${\cal T}$ is to convert a gauge invariant object $A$ into a new gauge invariant object ${\cal T} \cdot A$.  At the very minimum,  ${\cal T}$ should conform to the following baseline physical criteria:
\begin{itemize}
	\item[{\it i})] $\cal T$ preserves on-shell kinematics.
	\item[{\it ii})] $\cal T$ preserves gauge invariance.
\end{itemize}
Let us consider each of these conditions in turn.  


\subsubsection*{On-shell Kinematics}

A physical scattering amplitude of massless particles is well-defined on the support of the on-shell conditions,
$p_i p_i = p_i e_i =0$, and momentum conservation,
$\sum_i p_i = 0$. Representations of the amplitude that differ by terms which vanish on-shell are physically equivalent.  It is then crucial that our construction be agnostic to such differences.

To preserve the on-shell conditions, we simply define the physical scattering amplitude $A$ to be a function of $p_i p_j$, $p_i e_j$, and $e_i e_j$ for $i\neq j$.  For momentum conservation, we define the total momentum operator
\eq{
	{\cal P}_v \equiv \sum_i p_i v \, ,
}
where $i$ runs over all external legs and and $v$ labels any momentum or polarization vector.  Due to the implicit momentum-conserving delta function in $A$, we have that
\eq{ 
	{\cal P}_v \cdot A=0 \,.
}
To ensure that ${\cal T}$ conserves momentum it is then sufficient to require that
\eq{
	[\, {\cal P}_v \, , \, {\cal T} \,] \cdot A =0 \,,
	\label{eq:mom_con}
}
so ${\cal P}_v \cdot {\cal T} \cdot A ={\cal T} \cdot {\cal P}_v \cdot  A= 0$, {\it i.e.}~the transmuted amplitude also conserves momentum. This does not trivially hold for differential operators involving $p_i p_j$ and $p_i e_j$ as we will see explicitly in next section.
Since $[\, {\cal P}_v \, , \, {\cal P}_w \,]=0$, this operator itself conserves momentum.


\subsubsection*{Gauge Invariance}

A physical scattering amplitude should also be gauge invariant.  To incorporate this constraint we define a differential operator corresponding to the Ward identity on particle $i$,
\eq{
	{\cal W}_i \equiv  \sum_v p_i v \, \partial_{ v e_i} \, .
}
Here $v$ runs over all momentum and polarization vectors in the amplitude.  
Any gauge invariant amplitude is annihilated by this operator,
\eq{
	{\cal W}_i  \cdot A = 0 \, .
}
Perhaps unsurprisingly, ${\cal W}_i$ is itself momentum-conserving and gauge invariant.  To establish that ${\cal W}_i$ conserves momentum we compute the commutator
\begin{equation}
\begin{split}
&[\,  {\cal W}_i  \, , \,  {\cal P}_v \, ]= \sum_{j,w} [\, p_i w \, \partial_{w e_i}  \, , \, p_j v \, ]  
=   \delta_{v e_i} {\cal P}_{p_i} \, ,
\end{split}
\end{equation}
which vanishes when acting on a physical amplitude $A$, so $[\,  {\cal W}_i  \, , \,  {\cal P}_v \, ] \cdot A=0$.  Meanwhile, to verify that ${\cal W}_i$ is gauge invariant, we compute the commutator
\begin{equation}
\begin{split}
&[\, {\cal W}_i  \, , \, {\cal W}_j \, ]= \sum_{v,w} [\, p_i v \, \partial_{v e_i } \, , \, p_j w \, \partial_{w e_j } \, ]   
=0 \, ,
\end{split}
\end{equation}
which vanishes automatically.


\subsection{Transmutation Operators}

We are now equipped to derive a systematic basis of operators which preserve gauge invariance and on-shell kinematics.  To begin, we define an ansatz expressed in a basis of first order differentials,
\eq{
	{\cal T} \equiv  \sum\limits_{i,j } {\cal A}_{ij} \partial_{p_i p_j} + {\cal B}_{ij} \partial_{p_i e_j}  +{\cal C}_{ij} \partial_{e_i e_j}\, ,
}
where ${\cal A}_{ij}$, ${\cal B}_{ij}$, and ${\cal C}_{ij}$ are general functions of the external kinematic data.  For later notational convenience we choose ${\cal A}_{ii} = {\cal B}_{ii} = {\cal C}_{ii}=0$, which is consistent with the on-shell conditions. 
Let us now constrain this ansatz with  momentum conservation and gauge invariance.

First, requiring that ${\cal T}$ preserves conservation of momentum implies that
\begin{align}
[ \, {\cal T} \, , \, {\cal P}_v \, ] 
= \sum_{i,j ,k}   [ \, {\cal A}_{ij} \partial_{p_i p_j} + {\cal B}_{ij} \partial_{p_i e_j} \, , \, p_k v \, ] 
  =0 \, .
\end{align}
Choosing $v$ to be a momentum vector or polarization vector, respectively, we obtain
\eq{
	\sum_{i}  {\cal A}_{ij}  + {\cal A}_{ji} =\sum_{i }  {\cal B}_{ij}  = 0 \, ,
}
for all $j$.  So momentum conservation constrains every row and column of ${\cal A}_{ij}$ sums to zero, every column of ${\cal B}_{ij}$ sums to zero, and leaves ${\cal C}_{ij}$ unfixed.
This leads us to a simple basis of mutually commuting operators which span the solution set of these constraints, 
\begin{equation}
\begin{split}
{\cal T}_{ij} &\equiv \partial_{e_i e_j} \\
{\cal T}_{ijk} &\equiv \partial_{p_i e_j} - \partial_{p_k e_j} \\
{\cal T}_{ijkl} &\equiv \partial_{p_i p_j}-  \partial_{p_k p_j} +\partial_{p_k p_l}-  \partial_{p_i p_l} \, ,
\end{split}
\end{equation}
where all indices are distinct with the symmetry properties
\begin{equation}
\begin{split}
{\cal T}_{ij} &= {\cal T}_{ji} \\ 
{\cal T}_{ijk} &= -{\cal T}_{kji}  \\
{\cal T}_{ijkl} &= -{\cal T}_{kjil}= {\cal T}_{klij} =  -{\cal T}_{ilkj} \, .
\end{split}
\end{equation}
Thus we obtain a basis of momentum-conserving operators.


Second, we consider the gauge invariance of these basis operators.  A simple calculation yields
\begin{align} \label{eq:commutators} 
[\, {\cal T}_{ij} \, , \, {\cal W}_k \, ] &= 0 \nonumber \\
[\, {\cal T}_{ijk} \, , \, {\cal W}_l \, ] &=\delta_{il} {\cal T}_{ij}- \delta_{kl} {\cal T}_{jk} \\
[\, {\cal T}_{ijkl} \, , \, {\cal W}_m \, ] &= (\delta_{im}-\delta_{km}){\cal T}_{jml}  +(\delta_{jm}-\delta_{lm}) {\cal T}_{imk} \, , \nonumber
\end{align}
so ${\cal T}_{ij}$ is intrinsically gauge invariant while ${\cal T}_{ijk}$ and ${\cal T}_{ijkl}$ are not.  As we will later see, operators which are not intrinsically gauge invariant can still be {\it effectively} gauge invariant if the right-hand side of \Eq{eq:commutators} annihilates the amplitude.  

Note that ${\cal T}_{ijkl}$ is especially peculiar because it is a differential operator in $p_i p_j$, and so yields generic double poles when applied to a physical amplitude.  As a result, ${\cal T}_{ijkl}$ is not necessarily a generator of amplitudes, although we will see later that it can appear in certain soft limits.    In any case, hereafter we focus on ${\cal T}_{ij}$ and ${\cal T}_{ijk}$.  



\subsubsection*{Trace Operators}

Since ${\cal T}_{ij}$ is intrinsically gauge invariant we can apply it with impunity to any gauge invariant object to produce a new gauge invariant object.   As a differential operator,  ${\cal T}_{ij}$ eliminates all appearances of $e_i$ and $e_j$ except  when appearing in the combination $e_i e_j$. Hence  ${\cal T}_{ij}$ is equivalent to a dimensional reduction in which the polarization vectors $e_i$ and $e_j$ are chosen to be extra-dimensional, {\it i.e.}~orthogonal to the directions spanned by the external gluons.  This kinematic configuration sets $e_i e_j=1$ and $v e_i = v e_j=0$ for any momentum or polarization vector  $v$ associated with a gluon, which is mathematically equivalent to  applying ${\cal T}_{ij}$. From  this perspective it is obvious why ${\cal T}_{ij}$  conserves momentum while preserving gauge invariance on the remaining particles---it is simply an implementation of dimensional reduction. 

 At the level of scattering amplitudes, ${\cal T}_{ij}$ then transmutes a pair of gluons into a pair of scalars,
\eq{
	{\cal T}_{ij} \cdot  A(\cdots, g_i , g_j,\cdots) = A(\cdots, \phi_i \phi_j,\cdots) \, ,
}
where the ellipses denote spectator particles and the transmuted scalars should be interpreted as biadjoint scalars carrying the original color index as well as an additional dual color index.  
Here and throughout this paper, we use a notation where commas separate sets of particles which are ordered according to the dual color.  Since the original gluons do not carry dual color, each gluon in the original amplitude is separated by a comma.  In contrast, the transmuted biadjoint scalars carry a dual color ordering $i,j$ within a new dual color trace, so they are not separated by a comma.  Note that for the simple case of transmuting a pair of gluons, the ordering $i,j$ is irrelevant because ${\cal T}_{ij}$ is symmetric.  Since ${\cal T}_{ij}$ initializes a new dual color trace, we refer to it as a ``trace operator''.



\subsubsection*{Insertion Operators}

Unfortunately,  ${\cal T}_{ijk}$ is not intrinsically gauge invariant because it acts nontrivially on the momenta $p_i$ and $p_k$ generated by gauge transformations on $e_i$ and $e_k$. If, however, $e_i$ and $e_k$ have  already been eliminated prior to acting with ${\cal T}_{ijk}$, then there is no issue with gauge invariance.  In other words, ${\cal T}_{ijk}$ is effectively gauge invariant when appearing in tandem with enough supplemental transmutation operators.   For example, 
\eq{
	[\,   {\cal T}_{ik}  \cdot {\cal T}_{ijk}  \, , \, {\cal W}_l \, ] = \delta_{il} \, {\cal T}_{ik}\cdot  {\cal T}_{ij}- \delta_{kl} \, {\cal T}_{ik}\cdot  {\cal T}_{jk} \, ,
}
where the right-hand side vanishes on any physical amplitude which is necessarily multi-linear in $e_i$ and $e_k$.   Said another way, ${\cal T}_{ijk}$ is effectively gauge invariant if its commutator in \Eq{eq:commutators} annihilates the amplitude.  Putting this all together at the level of scattering amplitudes, we find that
\eq{
	{\cal T}_{ijk}  \cdot A(\cdots \phi_i \phi_k \cdots, g_j,\cdots)  
	=A(\cdots  \phi_i \phi_j \phi_k \cdots ,\cdots) \, .  \label{eq:insertion_claim}
}
As before, the transmuted states are biadjoint scalars carrying the original color as well as a new dual color.  Since the indices of ${\cal T}_{ijk}$ are ordered, these scalars have the ordering  $i,j,k$ within the dual color trace.  As noted earlier, the resulting biadjoint scalars are not separated by commas because they reside in the same dual color trace.
Because ${\cal T}_{ijk}$ transmutes a gluon into a biadjoint scalar and inserts it between two existing biadjoint scalars, we will refer to it as an ``insertion operator''.

\subsubsection*{Longitudinal Operators}

Let us now define a set of ``longitudinal operators'' which transmute particles into derivatively coupled longitudinal modes, {\it e.g.}~pions and Galileons.  These operators are defined as
\begin{equation}
{\cal L}_i \equiv \sum_j p_i p_j \partial_{p_j e_i}  \qquad \textrm{ and } \qquad
{\cal L}_{ij} \equiv - p_i p_j \partial_{e_i e_j}  \, , \label{eq:long_def}
\end{equation}
where $j$ runs over all momenta in the amplitude and we have included a relative sign for later convenience.  The longitudinal operators preserve on-shell kinematics because they are linear combinations of the trace and insertion operators: ${\cal L}_i = \sum_{j\neq k} p_i p_j {\cal T}_{jik}$ and ${\cal L}_{ij} = - p_i p_j {\cal T}_{ij}$ where $k$ is an arbitrary reference leg.  By the same logic, ${\cal L}_{ij}$ is intrinsically gauge invariant but ${\cal L}_{i}$ is not, since
\eq{
	[\, {\cal L}_i \, , \, {\cal W}_j \, ] &= -{\cal L}_{ij} \, . \label{eq:LW}
}
Just as before, however, ${\cal L}_i$ can still be effectively gauge invariant if the right-hand side of the commutator annihilates the amplitude.

It is clear from \Eq{eq:LW} that ${\cal L}_i$ and ${\cal L}_{ij} $ are closely related by gauge invariance.  We can exploit this kinship to derive a simple formula relating products of ${\cal L}_i$ to products of ${\cal L}_{ij}$.  To do this we write the Ward identity in the form
\eq{
	{\cal W}_i = {\cal L}_i + \Delta_i \, , \label{eq:Wi_to_Li}
}
where $\Delta_i = \sum_j p_i e_j \partial_{e_j e_i}$ is the difference between the Ward identity operator and the longitudinal operator.  Starting from any product of ${\cal L}_i$ operators, we insert ${\cal L}_i = {\cal W}_i - \Delta_i$ and commute factors of ${\cal W}_i$ to the right until they annihilate the amplitude.  Iterating this procedure, we find that
\begin{equation}
\begin{split}
	{\cal L}_i  \cdot {\cal L}_j &= {\cal L}_{ij} + \cdots  \\
	{\cal L}_i \cdot  {\cal L}_j  \cdot {\cal L}_k \cdot  {\cal L}_l &= {\cal L}_{ij}\cdot  { \cal L}_{kl}+{\cal L}_{ik} \cdot  {\cal L}_{jl}+{\cal L}_{il} \cdot  {\cal L}_{jk} + \cdots    \, ,\label{eq:Lrelation}
\end{split}
\end{equation} 
where the ellipses denote terms which have ${\cal W}_i$ commuted all the way to the right or which involve $\Delta_i$.    
In order to generalize \Eq{eq:Lrelation} we define a shorthand for a product of longitudinal operators,
\eq{
	{\cal L} \equiv \prod_{i} {\cal L}_{i} = \sum_{ \rho } \prod_{i,j\,\in \textrm{pairs}} {\cal L}_{ij} +\cdots, \label{eq:Ldef}
}
where in the first equality $i$ runs over all the gluons in the scattering amplitude and in the second equality $\rho$ runs over all partitions of the external gluons into pairs and $i,j$ run over each pair in a partition.  Again, the ellipses denote terms with ${\cal W}_i$ on the right of involving $\Delta_i$.  The former terms identically annihilate the physical amplitude while the latter terms vanish due to the multi-linearity of the amplitude in polarizations.

Since ${\cal L}_i$ and ${\cal L}_{ij}$ carry explicit factors of $p_i p_j$, they reduce the spin of states while also increasing the number of derivatives per particle.   
As we will see, these operators transmutes states into longitudinal modes, {\it e.g.}~pions or Galileons.  Note that ${\cal L}_{ij}$ is equivalent to the ``compactify'' operation for CHY integrands presented in \cite{Cachazo:2014xea}, here generalized to any representation of the amplitude.

\section{Unified Web of Theories}
\label{sec:web}

Applied in various combinations, the transmutation operators form an interlocking web of scattering amplitudes relations across a wide range of theories.   Remarkably, all S-matrices descend from the S-matrix of extended gravitons.
As discussed in the introduction, transmutation can be understood for YM and then straightforwardly generalized to extended gravity and BI via double copy construction.
Therefore, we focus here mainly on the transmutation on YM amplitudes.

Let us briefly review some of the basics of color structure in YM theory.
The full color-dressed YM amplitude can be expressed in terms of traces of color generators
\eq{
	A_{\rm YM}^{\rm full}(g_1, g_2,\cdots, g_n) = \sum_{\alpha \in S_n/Z_n} \textrm{Tr}(T^{\alpha_1}T^{\alpha_2}\cdots T^{\alpha_n}) \, A_{\rm YM}(g_{\alpha_1} g_{\alpha_2} \cdots g_{\alpha_n}),
	\label{eq:color_ordering}
}
where $\alpha$ sums over all orderings modulo cyclic permutations and the right-hand side is comprised of so-called color-ordered amplitudes.
As before, we use commas to delineate sets of particles with no relative ordering.  For this reason the states in the color-dressed amplitude in \Eq{eq:color_ordering} are separated by commas, while those in the color-ordered amplitudes are not.  Similarly, we would label a non-planar amplitude with commas separating sets of particles in difference traces.

This decomposition not only works for YM but also for the flavor group of pions. Also, in the case of BS theory, there are two color groups so we can apply the above color decomposition twice
\eq{
	A_{\rm BS}^{\rm full}(\phi_1, \phi_2,\cdots, \phi_n) = \sum_{\alpha,\beta \in S_n/Z_n} \textrm{Tr}(T^{\alpha_1}T^{\alpha_2}\cdots T^{\alpha_n}) \textrm{Tr}(T^{\beta_1}T^{\beta_2}\cdots T^{\beta_n})\, A_{\rm BS}(\phi_{\alpha_1} \cdots \phi_{\alpha_n} | \phi_{\beta_1} \cdots \phi_{\beta_n}) \nonumber
	\\
	\label{eq:double_color}
}
where $A_{\rm BS}(\phi_{\alpha_1} \cdots \phi_{\alpha_n} | \phi_{\beta_1} \cdots \phi_{\beta_n}) $ is the doubly color-ordered amplitude encoding a color and dual color ordering, often abbreviated as $A_{\rm BS}(\alpha|\beta)$. For notational simplicity, we will often keep the dual color ordering arising from transmutation explicit, while suppressing the original color ordering inherited from starting amplitude.

We stress here that the insertion operators, which transmute graviton to gluons, gluons to biadjoint scalars, or BI photons to pions, produce color-ordered amplitudes.
The purpose of insertion in this paper is to reduce the spins but project to an ordering of the particles, which we will show concretely in Sec.~\ref{sec:BCJ}.
The mapping of the full amplitudes, however, requires summing all orderings dressed with color factors put by hands.
Let us discuss the resulting single trace and multiple trace scattering amplitudes from transmutation in turn.

\subsection{Single Trace Amplitudes}

To begin, it will be convenient to define a shorthand notation denoting a single trace operator followed by a sequence of insertion operators,
\eq{
	{\cal T}[\alpha] \equiv    {\cal T}_{\alpha_1 \alpha_n} \cdot   \prod_{i =2}^{n-1} {\cal T}_{\alpha_{i-1} \alpha_i \alpha_n}  \, ,   \label{eq:Odef}
}
where $\alpha$ is an ordered set.  
Applying ${\cal T}[\alpha] $ to a scattering amplitude of gluons, we obtain
\begin{equation}
{\cal T}[i_1 \cdots i_n]  \cdot A(g_{i_1},\cdots ,g_{i_n},\cdots) = A( \phi_{i_1} \cdots \phi_{i_n},\cdots) \, ,
\label{eq:master}
\end{equation}
which is an amplitude of biadjoint scalars coupled to the remaining nontransmuted gluons denoted by the ellipses after the last comma.
If $\alpha$ is the set of  all gluons, then no gluons remain after transmutation and the resulting amplitude is that of the BS theory.  If, on the other hand, $\alpha$ is a subset of gluons, then we obtain an amplitude of the gauged BS theory, which describes biadjoint scalars interacting with gluons through the original color index.  Since the dual color ordering $\alpha$ of the biadjoint scalars is defined modulo cyclic permutations, ${\cal T}[\alpha]$ is actually cyclically invariant, though not manifestly so.

Remarkably, the above construction is quite general. For example,  by transmuting a extended gravity amplitude, we obtain
\eq{
	{\cal T} [i_1 \cdots i_n]  \cdot  A(h_{i_1},\cdots ,h_{i_n},\cdots) =A(g_{i_1} \cdots g_{i_n}, \cdots) \, ,
}
which is an amplitude of gluons coupled to extended gravitons, {\it i.e.}~an EYM amplitude.  Applied to photon scattering amplitudes in BI theory, transmutation yields
\eq{
	{\cal T}[i_1 \cdots i_n]  \cdot A(\gamma_{i_1}, \cdots, \gamma_{i_n},\cdots) = A( \pi_{i_1} \cdots \pi_{i_n}, \cdots) \, ,
}
which is an amplitude of pions coupled to BI photons.

Meanwhile, the longitudinal operator transmutes all the gluons in an amplitude to pions,
\eq{
	{\cal L} \cdot A(g_{i_1}, \cdots, g_{i_n}, \cdots)  
	=A(\pi_{i_1}, \cdots, \pi_{i_n}, \cdots) \, ,  \label{eq:long_claim}
}
where the ellipses after the last comma denote any spectator biadjoint scalars in the amplitude.
If the initial amplitude does not contain any biadjoint scalars, then the transmuted object actually vanishes.  
If, on the other hand, the initial amplitude starts with  precisely two biadjoint scalars, then the resulting amplitude is actually equal to an amplitude of all external pions.
  That is, irrespective of which pair of external legs are initially chosen to be biadjoint scalars, the longitudinal operator produces a permutation invariant amplitude of the NLSM.  This obfuscation of permutation invariance is reminiscent of the cubic NLSM action proposed in \cite{Cheung:2016prv}.  Finally, if the initial amplitude has three or more biadjoint scalars, then the resulting amplitude is that of the extended NLSM \cite{Cachazo:2016njl}, describing pions coupled to biadjoint scalars.

As before, this construction also applies to other theories.  In particular, the longitudinal operator transmutes an amplitude of extended gravitons coupled to gluons  into
\begin{equation}
	{\cal L} \cdot A(h_{i_1}, \cdots , h_{i_n}, \cdots)  
	=A(\gamma_{i_1} , \cdots , \gamma_{i_n}, \cdots) \, ,
\end{equation}
which is an amplitude of BI photons coupled to gluons.
Similarly, the longitudinal operator transmutes an amplitude of BI photons coupled to pions into
\begin{equation}
{\cal L} \cdot  A(\gamma_{i_1}, \cdots , \gamma_{i_n}, \cdots)
= A(\phi_{i_1}, \cdots, \phi_{i_n}, \cdots) \, ,
\end{equation}
which is an amplitude of SG scalars coupled to pions.

\subsection{Multiple Trace Amplitudes}

It is straightforward to generalize our results to include multiple trace structures.  To start, let us consider the simplest multiple trace amplitudes where each trace is comprised of a pair of states.  Starting from a gluon amplitude, we obtain
\eq{
	{\cal T}[i_1 j_1] \cdots {\cal T}[i_m j_m] \cdot   A(g_{i_1} , g_{j_1} , \cdots  , g_{i_m} , g_{j_m},\cdots)  
= A(\phi_{i_1 }\phi_{j_1}, \cdots , \phi_{i_m }\phi_{j_m},\cdots) \, , \label{eq:trace_claim} 
}
where the ellipses after the last comma again denote nontransmuted spectator gluons.
The resulting amplitude describes biadjoint scalars coupled to gluons via gauge interactions.
This operation also transmutes extended gravity into EM theory and BI theory into the DBI scalar theory.

To discuss more complex trace structures it will be convenient to make the  dependence on the original color ordering explicit. In particular, if we start with a color ordered gluon amplitude $A(\beta)$, then \Eq{eq:master} can be written in the shorthand
\eq{
	{\cal T}[\alpha] \cdot A(\beta)  = A( \alpha |\beta )  \, ,
}
in the notation of \Eq{eq:Odef}. The $A( \alpha |\beta )$ on the right-hand side is the doubly color-ordered amplitude for BS theory in \Eq{eq:double_color}.
Applying this operator multiple times, we obtain
\eq{
	{\cal T}[\alpha_1]  \cdots {\cal T}[\alpha_m] \cdot A(\beta) &= A(\alpha_1, \cdots,   \alpha_m  |\beta) \, ,
}
which is a multiple trace amplitude in gauged BS theory.

In contrast, since the amplitudes of extended gravity and BI theory are not color-ordered to start,
\eq{
	{\cal T}[\alpha] \cdot A  = A( \alpha) \,.
}
In this case the multiple trace amplitude is
\eq{
	{\cal T}[\alpha_1]  \cdots {\cal T}[\alpha_m]  \cdot A &= A(\alpha_1, \cdots, \alpha_m ) \, ,
}
corresponding to the amplitudes of the NLSM coupled to BI photons and EYM theory.

\subsection{Summary of Unifying Relations}

Let us briefly summarize the unifying relations associated with some better-known theories, corresponding to the corners of the web depicted in \Fig{fig:unified_web}.  From YM amplitudes we obtain,
\begin{equation}
\begin{split}
	A_{\rm YMS} &= {\cal T}[i_1 j_1]\cdots  {\cal T}[ i_m j_m]  \cdot  A_{\rm YM}  \\
	A_{\rm BS} &= {\cal T}[i_1 \cdots i_n]\cdot A_{\rm YM}  \\
	A_{\rm NLSM} &={\cal L} \cdot {\cal T}[i_1 i_n] \cdot A_{\rm YM} \, .
\end{split}
 \label{eq:YM_sum}
\end{equation}
From BI amplitudes we obtain
\begin{equation}
\begin{split}
	A_{\rm DBI} &= {\cal T}[i_1 j_1]\cdots  {\cal T}[ i_m j_m]  \cdot  A_{\rm BI}  \\
	A_{\rm NLSM} &= {\cal T}[i_1 \cdots i_n]\cdot A_{\rm BI}  \\
	A_{\rm SG} &={\cal L} \cdot {\cal T}[i_1 i_n] \cdot A_{\rm BI} \,.
\end{split}
	\label{eq:BI_sum}
\end{equation}
From extended gravity amplitudes we obtain
\begin{equation}
\begin{split}
	A_{\rm EM} &= {\cal T}[i_1 j_1]\cdots  {\cal T}[ i_m j_m]  \cdot  A_{\rm G}  \\
	A_{\rm YM} &= {\cal T}[i_1 \cdots i_n]\cdot A_{\rm G}  \\
	A_{\rm BI} &={\cal L} \cdot {\cal T}[i_1 i_n] \cdot A_{\rm G} \,. 
\end{split}
\label{eq:GR_sum}
\end{equation}
Since the above amplitudes are all ultimately derived from extended gravity, they share the kindred relations depicted in \Fig{fig:unified_web}.    

A trivial corollary of these unifying relations is the universality of the KLT, BCJ, and CHY constructions.  Because these structures generalize to arbitrary spacetime dimension, the associated scattering amplitudes can be represented as functions of Lorentz invariant products of momentum and polarization vectors---so transmutation applies.  Given YM theory and gravity as input,  it is then straightforward to import the KLT and BCJ relations into all the theories in \Fig{fig:unified_web} via transmutation.  Similarly, the CHY integrand for gravity can be transmuted into new CHY integrands for the descendant theories.  We leave a full analysis of these ideas for future work.

\subsection{Ultraviolet Completion}
\label{sec:more}

Transmutation is based on the dual assumptions of gauge invariance and on-shell kinematics---for massless external particles.  Notably, these stipulations do not preclude the presence of massive particles exchanged as intermediate states, as would arise from ultraviolet completion.    Hence, transmutation also defines a web of unifying relations for the {\it ultraviolet completions} of the theories in \Fig{fig:unified_web}.  Whether the resulting amplitudes are fully consistent is not guaranteed---after all, our factorization proof assumes a massless spectrum---but it is still illuminating to see the variety of objects that arise from this procedure.

To begin, consider an ultraviolet completion of YM theory in which new massive degrees of freedom arrive at a physical scale suggestively denoted by $\alpha'$.     Since the amplitude is a gauge invariant object, transmutation generates a new gauge invariant object,
\eq{ 
	{\cal T}[i_1 \cdots i_n] \cdot A_{\rm YM+ \alpha'}(g_{i_1}, \cdots, g_{i_n}) = A_{\rm BS + \alpha'}(\phi_{i_1} \cdots \phi_{i_n} )  \, ,
}
corresponding to a possible ultraviolet completion of the BS theory with $\alpha'$ corrections. 
A natural candidate for an ultraviolet completion of YM theory is open superstring theory, $A_{\rm YM + \alpha'} \sim A_{\rm open}$, in which case the transmutation equation becomes
\eq{  \label{eq:stringrelation1}
	{\cal T}[i_1 \cdots i_n] \cdot A_{\rm open}(g_{i_1}, \cdots, g_{i_n}) = A_{\rm Z}(\phi_{i_1} \cdots \phi_{i_n} )  \, ,
}
where $ A_{\rm BS + \alpha'} \sim A_{\rm Z}$  is the amplitude of biadjoint scalars obtained from the Z-theories of Carrasco, Mafra, and Schlotterer \cite{Carrasco:2016ldy, Mafra:2016mcc, Carrasco:2016ygv}. Explicitly, this amplitude is equal to the world sheet integral
\eq{
	A_{\rm Z}(\phi_{i_1} \cdots \phi_{i_n})
	= 
	(\alpha')^{n-3} \int_{D  } \, { dz_1 dz_2 \cdots dz_n \over {\rm vol} (SL(2,R)) } 
	{\prod^n_{i<j} |z_{ij}|^{\alpha' p_i p_j} \over z_{i_1 i_2} z_{i_2 i_3} \cdots  
		z_{i_n i_1} } \,, 
}
where $z_{ij} = z_i - z_j$, and the integration domain $D$ is along the real line, 
\eq{
	D = \{  - \infty < z_1 < z_2< \cdots <z_n < + \infty\} \, .
}
An analogous argument for the ultraviolet completions of BI theory and the NLSM imply that
\eq{ \label{eq:stringrelation2}
	{\cal T}[i_1 \cdots i_n] \cdot A_{\rm open}( \gamma_{i_1}, \ldots , \gamma_{i_n}) = A_{ \rm Z }( \pi_{i_1} \cdots \pi_{i_n} )  \, ,
}
corresponding to the BI photon amplitudes of open superstring theory and the pion amplitudes of abelian Z-theory \cite{Carrasco:2016ldy}, respectively.

The relations in \Eq{eq:stringrelation1} and \Eq{eq:stringrelation2} can be trivially derived from the KLT relations.  Based on earlier results in \cite{Mafra:2011nv, Mafra:2011nw}, the open superstring amplitudes can be recast as a KLT product of amplitudes in Z-theory and YM theory \cite{Broedel:2013tta},   
\eq{ \label{eq:string1}
	A_{\rm open} =  A_{\rm Z } \otimes A_{\rm YM}\, ,
}
where $\otimes$ denotes the field theory KLT product. 
Transmuting the left-hand side sends $A_{\rm YM}$ to $A_{\rm BS}$ on the right-hand side. Since a KLT product with $A_{\rm BS}$ is trivial, we obtain ${\cal T}[i_1 \cdots i_n] \cdot	A_{\rm open} =  A_{\rm Z } \otimes A_{\rm BS} = A_{\rm Z} $, which is equivalent to \Eq{eq:stringrelation1} and \Eq{eq:stringrelation2}.  Instead applying a longitudinal operator, we obtain  $ \mathcal{L} \cdot \mathcal{T}[i_1 i_n ]\cdot	A_{\rm open}  =  A_{\rm Z} \otimes A_{\rm NLSM}$, whose scalar amplitudes coincide with the NLSM at low energies, but curiously does not correspond to abelian Z-theory.

A similar logic also holds for extended gravity and its ultraviolet completion.   Transmuting the graviton scattering amplitude yields
\eq{ \label{eq:stringrelation3}
	{\cal T}[i_1 \cdots i_n] \cdot {A}_{\rm G+ \alpha'}(h_{i_1}, \cdots, h_{i_n}) =
	A_{\rm YM + \alpha' } (g_{i_1}\cdots g_{i_n}) \, .
}
Interestingly, when taking $A_{\rm G + \alpha'} \sim A_{\rm closed}$ to be amplitudes of the closed superstring, we find that the right-hand side corresponds not to open superstring amplitudes, but to the so-called single value projection  \cite{Brown:2013gia} \cite{Schlotterer:2012ny, Stieberger:2013wea, Stieberger:2014hba} of the open-string amplitudes, $A_{\rm YM + \alpha'} \sim A_\textrm{open}^{\rm SV}$.  For a detailed discussion on the single-value projection, see for instance \cite{Stieberger:2014hba} and references there in.
In any case, the above claim is understood by the convenient KLT form for closed superstring amplitudes 
\eq{ \label{eq:closestringKLT2}
	{A}_{\rm closed }  = 
	A_\textrm{open }^{\rm SV}  \otimes
A_{\rm YM} \,,
}
where the KLT kernel is for field theory rather than for string theory \cite{Kawai:1985xq}.


\section{Examples}
\label{sec:example}

To familiarize the reader with the mechanics of transmutation, let us analyze a few concrete examples.  Later on, some of these will serve as base cases for a proof by induction of our claims.

\subsection{Three-Particle Amplitudes}

To begin, consider the three-particle scattering amplitude of gluons,
\eq{
	A(g_1 ,g_2, g_3) = \frac{1}{2} e_1   e_2 (p_2 e_3 - p_1 e_3) + \textrm{cyclic} \, ,
}
where the YM color ordering $1,2,3$ is implicitly assumed and only dual color ordering is shown.  By applying the trace operator, we  transmute any two gluons into biadjoint scalars,
\begin{equation}
\begin{split}
A(\phi_1 \phi_2, g_3) &=	{\cal T}[12]\cdot A(g_1, g_2, g_3) = \frac{1}{2} (p_2 e_3 - p_1 e_3)    \\
 A(\phi_2 \phi_3, g_1) &=	{\cal T}[23]\cdot  A(g_1, g_2, g_3) =  \frac{1}{2}(p_3 e_1 - p_2 e_1)  \\
A(\phi_3 \phi_1, g_2) &= 	{\cal T}[31]\cdot  A(g_1, g_2, g_3) =  \frac{1}{2}(p_1 e_2 - p_3 e_2)  \,  . 
\end{split}
\label{eq:A3example}
\end{equation}
Alternatively, we can transmute all three gluons simultaneously,
\eq{
A(\phi_1 \phi_2 \phi_3) &=	{\cal T}[123]\cdot  A(g_1 , g_2, g_3) = 1 \, ,
\label{eq:phi3amp}
}
into biadjoint scalars with the dual color ordering $1,2,3$ here.  Here we have dropped the dimensionful cubic coupling associated with the BS theory.  
For the reverse ordering $3,2,1$, we apply
\eq{
A(\phi_3 \phi_2 \phi_1) &=	{\cal T}[321]\cdot  A(g_1, g_2, g_3) = -1 \, .
}
The relative minus sign for the second case captures the anti-symmetry of the dual color factor. 

Finally, applying the longitudinal operator to \Eq{eq:A3example} yields
\begin{equation}
\begin{split}
A(\pi_1, \pi_2, \pi_3) &= A(\phi_1  \phi_2, \pi_3) = {\cal L}\cdot	{\cal T}[12]\cdot A(g_1, g_2, g_3) \\
&= A(\phi_2  \phi_3, \pi_1) = {\cal L}\cdot	{\cal T}[23]\cdot A(g_1, g_2, g_3) \\
&= A(\phi_3  \phi_1, \pi_2) = {\cal L}\cdot	{\cal T}[31]\cdot A(g_1, g_2, g_3) =  0 \, ,
\end{split}
\end{equation}
where as noted earlier, we can make any choice for the pair of biadjoint scalars to obtain the permutation invariant pion amplitude.  As expected, the three-particle amplitude for pions is zero.

\subsection{Four-Particle Amplitudes}

Next, let us apply transmutation to the four-particle scattering amplitude of gluons.  The expression from color-ordered Feynman diagrams yields\footnote{Our normalization convention for amplitudes effectively sets all propagator denominators to $\sum_{i\neq j} p_i p_j$.}
\eq{
A(g_1,g_2,g_3,g_4) &= \frac{n_{[12][34]}}{p_1 p_2}+\frac{n_{[23][41]}}{p_2 p_3}
-\frac{1}{2}\left[ (\ee{1}{2}) (\ee{3}{4})+(\ee{2}{3}) (\ee{4}{1}) \right]
+ (\ee{1}{3}) (\ee{2}{4}) \, ,
}
where the numerator of the cubic diagram is $n_{[12][34]} = n_{1234} - n_{2134} - n_{1243} + n_{2143} $ and
\eq{
	n_{1234} =& 
\frac{1}{4} (\ee{1}{2}) (\ee{3}{4}) (\pp{1}{3})
	-(\ee{2}{3}) (\pe{3}{4}) (\pe{2}{1})
	-\frac{1}{2} (\ee{1}{2}) (\pe{2}{3})(\pe{1}{4})
	-\frac{1}{2} (\ee{3}{4}) (\pe{4}{1}) (\pe{3}{2}) \, ,  
}
and similarly for $n_{[23][41]}$ up to relabelling. Again, the YM color ordering $1,2,3,4$ is being suppressed and we only display the ordering of the dual color group.
By transmuting two pairs of gluons, we obtain the scattering amplitudes for biadjoint scalars,
\begin{equation}
\begin{split}
A(\phi_1 \phi_2, \phi_3 \phi_4) &= {\cal T}[12]\cdot {\cal T}[34]\cdot A(g_1,g_2,g_3,g_4) =  \frac{p_1 p_3}{p_1 p_2}  \\
A(\phi_1 \phi_3, \phi_2 \phi_4) &= {\cal T}[13]\cdot {\cal T}[24]\cdot A(g_1,g_2,g_3,g_4) =  1 \, .
\end{split}
\end{equation}
Note the presence of a contact quartic scalar interaction in both cases.

Meanwhile, the transmuted single trace four-particle amplitudes are
\begin{equation}
\begin{split}
A(\phi_1 \phi_2, g_3, g_4) & = {\cal T}[12]\cdot A(g_1, g_2, g_3, g_4)  = 
\frac{(p_1 e_3) (p_2 e_4) + (p_1 p_3) (e_3 e_4) - (p_1 e_4) (p_2 e_3) }{p_1 p_2}
-\frac{ (p_1 e_4) (p_2 e_3) }{p_2 p_3} \\ 
A(\phi_1 \phi_2 \phi_3, g_4) & = {\cal T}[123]\cdot A(g_1, g_2 ,g_3, g_4) =\frac{p_3 e_4}{p_1 p_2}-\frac{p_1 e_4}{p_1 p_4} \\
A(\phi_1 \phi_2 \phi_3 \phi_4) & = {\cal T}[1234]\cdot A(g_1, g_2, g_3, g_4) = \frac{1}{p_1 p_2} + \frac{1}{p_2 p_3} \, .
\end{split} 
\label{eq:4pt_example}
\end{equation}
Note that the ordering in the transmutation operator is crucial because it dictates the order in the dual color trace.  In in particular, the relative ordering of the original and dual color traces will affect the resulting scattering amplitude.  So for example, we find that
\begin{equation}
\begin{split}
A(\phi_1 \phi_3 \phi_2 \phi_4) &=	{\cal T}[1324]\cdot  A(g_1, g_2, g_3, g_4) = -\frac{1}{p_2 p_3} \\
 A(\phi_1 \phi_3 \phi_4 \phi_2) &= 	{\cal T}[1342]\cdot  A(g_1, g_2, g_3, g_4)= -\frac{1}{p_1 p_2} \, ,
\end{split}
\end{equation}
so as expected, these amplitudes exhibit factorization channels for particles which are adjacent in both the original color and the dual color. Furthermore, we have,
\eq{
A(\phi_1 \phi_2 \phi_3 \phi_4) + A(\phi_1 \phi_3 \phi_2 \phi_4)  + A(\phi_1 \phi_3 \phi_4 \phi_2) = 0 \, ,
}
which is the transmuted form of the $U(1)$ decoupling relation.

Lastly, let us consider transmutation via the longitudinal operators. Again applying the freedom of choosing an initial pair of biadjoint scalars, we obtain
\begin{equation}
\begin{split}
A(\pi_1, \pi_2, \pi_3, \pi_4) &= A(\phi_1 \phi_2, \pi_3, \pi_4) = 
{\cal L}\cdot {\cal T}[12]\cdot A(g_1, g_2, g_3, g_4) \\
&= A(\phi_1 \phi_3, \pi_2, \pi_4) =  {\cal L} \cdot {\cal T}[13]\cdot A(g_1, g_2, g_3, g_4) = p_1 p_3  \, , 
\end{split}
\end{equation}
which is the correct four-particle color-ordered pion amplitude. 


\subsection{Gluon and Graviton Amplitudes}
\label{sec:BCJ}

Last but not least, consider amplitudes comprised of all gluons or all extended gravitons. We start by proving \Eq{eq:master} in the special case where all the external states are gluons. 
A tree amplitude in YM can be expressed as,
\eq{
	A^{\rm full}_{\rm YM} = \sum_i \frac{N_i C_i}{D_i} \, ,
}
where $C_i$ are the color factors and $N_i$ are respectively the color and kinematic BCJ numerators associated with a cubic graph $i$. 
The color and kinematic numerators can be expanded in the Del Duca-Dixon-Maltoni half-ladder bases \cite{DelDuca:1999rs},
\eq{
	N_i = \sum_\alpha \sigma_{i}(\alpha) N(\alpha) \qquad \textrm{and} \qquad   C_i = \sum_\beta  \sigma_{i}(\beta) C(\beta) \, .
}
Projecting the full amplitude into the color ordering $\beta$,
\eq{
	A_{\rm YM}(\beta) = \sum_\alpha  N(\alpha) A_{\rm BS}(\alpha |\beta) \, ,
}
where $A_{\rm BS}(\alpha | \beta) $ is actually the double-trace biadjoint scalar amplitude
\eq{
	A_{\rm BS}(\alpha | \beta) =  \sum_i \frac{\sigma_{i}(\alpha) \sigma_{i}(\beta) }{D_i} \, ,
}
which forms a basis of  doubly color-ordered biadjoint scalar amplitudes.
As we will show shortly, the operator in \Eq{eq:Odef} is constructed so that
\eq{
	{\cal T}[\alpha] \cdot N(\beta) = \delta_{\alpha \beta} \, , \label{eq:inverse}
}
which is one if $\alpha$ and $\beta$ are the same permutation and zero otherwise.  Applying this to the color-ordered gluon amplitude yields precisely \Eq{eq:master}.
Thus, this operator extracts precisely  the doubly color-ordered biadjoint scalar amplitude from the color-ordered YM amplitude.

Meanwhile, we can lift this construction to extended gravity by again utilizing the double copy,
\eq{
	A_{\rm G} = \sum_i \frac{\tilde N_i N_i }{D_i}  = \sum_{\alpha, \beta} \tilde N(\alpha) A_{\rm BS}(\alpha|\beta) 
	N(\beta) = \sum_{ \beta} N(\beta) A_{\rm YM}(\beta) \, .
}
Using the relation in \Eq{eq:inverse} we find that 
\eq{
	{\cal T}[\alpha] \cdot A_{\rm G} =  A_{\rm YM}(\alpha) \, , \label{eq:GRtoYM}
}
so we obtain color-ordered gluon amplitudes from graviton amplitudes.

To derive \Eq{eq:inverse} we will use the fact that the BCJ numerator for the gluon amplitude takes the schematic form
\eq{
	N(\beta) \sim (e e)(pe)^{n-2} + (e e)^2(pe)^{n-4}(pp) +  \ldots \, ,
	\label{eq:schematic}
}
where the ellipses denote terms with higher and higher powers of $(ee)$.  The above schematic expression of $N(\beta)$ is determined simply by the power counting. By construction, the operator ${\cal T}[\alpha]$ eliminates a single factor of $(ee)$ and $n-2$ factors of $(pe)$, so it projects out all but the first term.   Because this term carries no powers of $(pp)$, this term in the numerator can never cancel propagator poles, so it must correspond to the pure cubic Feynman diagram contributions to the gluon amplitude.  It is trivial to compute the cubic Feynman diagram contribution, yielding
\eq{
	N(\alpha) = e_{\beta_1} e_{\beta_n} \prod_{i=2}^{n-1} \sum_{j=1}^{i-1} p_{\beta_j} e_{\beta_i} + \ldots \, ,
} 
where the ellipses denote numerous other permutations of terms that are all annihilated by ${\cal T}[\alpha]$. It finishes the proof of Eq. (\ref{eq:inverse}). Finally we remark that Eq. (\ref{eq:GRtoYM}) may also be argued applying the statement of \cite{Arkani-Hamed:2016rak}, which states that scattering amplitudes in YM theories are uniquely fixed by the requirement of simple poles, power counting and gauge invariance, and the operator ${\cal T}[\alpha]$ clearly preserves all those properties.

\section{Proof by Induction}
\label{sec:proof}

We now present an induction proof for the unifying relations summarized in \Sec{sec:web} and depicted diagrammatically in \Fig{fig:unified_web}.  
In particular, we show that  ${\cal T}_{ij}$, ${\cal T}_{ijk}$, and ${\cal L}$
 transmute physical amplitudes into other physical amplitudes.   In our earlier discussion, we observed that ${\cal T}_{ijk}$ and ${\cal L}$ are only gauge invariant when applied to an amplitude already containing states transmuted by ${\cal T}_{ij}$.   For this reason we must initialize the induction with a base case: the amplitude in \Eq{eq:trace_claim}, which describes any number of pairs of biadjoint scalars in distinct dual color traces interacting with gluons. 
 Any single or multiple trace amplitude is obtained by applying ${\cal T}_{ijk}$ and ${\cal L}$ to this amplitude.   Other useful base cases are the three- and four-particle amplitudes derived in \Sec{sec:example}.


In order to establish the unifying relations we demonstrate that a transmuted amplitude satisfies necessary and sufficient conditions that define a physical amplitude.  This proof occurs in three steps.  First, 
we verify that the transmuted amplitude is on-shell constructible, {\it i.e.} determined by the residues on each factorization channel and hence defined via on-shell recursion relations.  As we will see, it will suffice that the amplitude  vanish either at large momentum transfer or in the soft limit.  Second, we check explicitly that the transmuted amplitude factorizes properly into products of physical amplitudes, using crucially that the same is true of the original amplitude by the induction hypothesis.
In this way we prove the unifying relations for single and multiple trace scattering amplitudes of gluons, biadjoint scalars, and pions summarized in \Eq{eq:YM_sum} and portrayed collectively by the lower triangle in Fig.~\ref{fig:unified_web}.  Third, we use the double copy construction to extend our proof  to the remaining unifying relations listed in \Eq{eq:BI_sum} and \Eq{eq:GR_sum} and depicted in \Fig{fig:unified_web}.


\subsection{On-shell Constructibility}

Consider an initial amplitude $A$ comprised of gluons and biadjoint scalars.  Using the results of \cite{Britto:2004ap,Britto:2005fq,Cohen:2010mi,Cheung:2015cba}, $A$ is is on-shell constructible by an all-line recursion relation.  In particular, for greater than four particles, the deformed amplitude $A$ falls off at large $z$ for an all-line momentum shift parameterized by $z$.  This result follows straightforwardly from dimensional analysis and does not require detailed information about the amplitude \cite{Cohen:2010mi,Cheung:2015cba}.

Applying the insertion operator to $A$ yields a new object ${\cal T}_{ijk} \cdot A$.   However, since ${\cal T}_{ijk} \cdot A$ carries strictly fewer derivatives than $A$, we can apply the exact same power counting argument from  \cite{Cohen:2010mi,Cheung:2015cba} to establish vanishing large $z$ behavior of ${\cal T}_{ijk} \cdot A$.  Thus, we determine that this object is also on-shell constructible.

Instead applying the longitudinal operator, we obtain a transmuted object ${\cal L}\cdot A$ with more derivatives than $A$, so large $z$ fall-off is not guaranteed.  Fortunately, an amplitude without large $z$ fall-off can still be on-shell constructible if it has compensating infrared properties like the Adler zero or its generalizations \cite{Cheung:2015ota,Luo:2015tat}.  As we will show in \Sec{sec:soft}, the pions in ${\cal L} \cdot A$ indeed manifest the Adler zero as a transmutation of the Weinberg soft theorems of gauge theory.   This applies even in the presence of spectator biadjoint scalars.  In particular, the growth in large $z$ behavior incurred from the derivatives in ${\cal L}$ is always balanced by the Adler zeros arising from the resulting pions.


\subsection{Factorization}

Since the transmuted amplitude has either large $z$ fall-off or vanishing soft limits, our remaining task is to verify proper factorization.  By the assumption of the induction hypothesis, the original initial $A$ factorizes according to
\eq{
	A(\cdots) &\sim \sum_{I } A_L(\cdots I_L)  \, A_R(I_R \cdots) \, ,
}
where $\sim$ denotes the residue on the factorization channel of the singular propagator, $1/p_L^2 = 1/p_R^2$.  Here the sum on $I$ runs over all possible gluon and scalar internal states.  In what follows, we certify the proper factorization of  ${\cal T}_{ijk} \cdot A$ and ${\cal L} \cdot A$, respectively.

\subsubsection*{Insertion Operator}

As shown in \Eq{eq:insertion_claim}, the insertion operator ${\cal T}_{ijk}$ transmutes the gluon $g_j$ into a biadjoint scalar, inserting  it between the two biadjoint scalars $\phi_i$ and $\phi_k$ which are adjacent in the dual color.  There are a handful of possible arrangements of these states relative to the factorization channel.  Let us consider each case in turn, demonstrating that the resulting amplitude factorizes correctly.

\medskip

\noindent {\it Gluon on Same Side as Scalars.}
If the gluon and both biadjoint scalars are on the same side of the factorization channel, then initial scattering amplitude factorizes according to 
\begin{equation}
A(\cdots\phi_i \phi_k\cdots, g_j, \cdots) \sim \sum_I A_L(\cdots\phi_i \phi_k\cdots, g_j,\cdots, I_L) \, A_R(I_R,\cdots) \, .
\end{equation}
Applying an insertion operator to both sides, we otain
\begin{equation}
\begin{split}
&{\cal T}_{ijk} \cdot A(\cdots\phi_i \phi_k\cdots, g_j, \cdots) \sim \sum_I A_L(\cdots\phi_i \phi_j \phi_k\cdots, \cdots, I_L) \,  A_R(I_R, \cdots) \, ,
\end{split}
\end{equation}
where in the second line we have used the induction hypothesis, which is that the claim in \Eq{eq:master} holds for all lower point amplitudes.  Hence, in this case the left-hand side factorizes correctly, {\it i.e.}~as the transmuted $A(\cdots \phi_i \phi_j\phi_k \cdots, \cdots)$ should.

\medskip

\noindent {\it Gluon on Opposite Side as Scalars.}
If the gluon is on the opposite side of the factorization channel from the biadjoint scalars, then 
\begin{equation}
\begin{split}
& A(\cdots\phi_i \phi_k\cdots ,  g_j, \cdots) \sim \sum_I A_L(\cdots\phi_i \phi_k\cdots, I_L)  \, A_R(I_R, g_j \cdots) \\
&\sim A_L(\cdots\phi_i \phi_k\cdots \phi_L)  \, A_R(\phi_R,  g_j \cdots) +A_L(\cdots\phi_i \phi_k\cdots , g_L)  \, A_R(g_R,  g_j \cdots) \,  , 
\end{split}
\end{equation}
where the first and second terms correspond to intermediate scalar and gluon exchange, respectively.  Next, consider ${\cal T}_{ijk}$ applied to each of these terms.

The first term is annihilated by ${\cal T}_{ijk}$  because the gluon polarization $e_j$ can only dot into the combination $p_i + p_k$ because the intermediate particle is a scalar and cannot transmit differences of momenta across the channel. This agrees with the factorization of $A(\cdots \phi_i \phi_j \phi_k\cdots, \cdots)$, which should vanish on this factorization channel since it requires $\phi_j$ to be adjacent to other external scalars in $A_R$ but also to be inserted between $\phi_i$ and $\phi_k$, which is impossible.

The second term is not annihilated by ${\cal T}_{ijk}$, but since the intermediate particle is a gluon we can apply the completeness relation,
\eq{
	\partial_{p_i e_j } - \partial_{p_k e_j}  \rightarrow    (\partial_{p_i e_{L} } - \partial_{p_k e_{L}}) \partial_{e_R e_{j}} \, ,
}
which expressed in terms of transmutation operators is
\eq{
	{\cal T}_{ijk} \rightarrow {\cal T}_{iLk}  {\cal T}_{Rj} \, .
}
Applying this to the factorization equation, we obtain
\begin{equation}
{\cal T}_{ijk} \cdot A(\cdots\phi_i \phi_k\cdots, g_j,\cdots) \sim A_L(\cdots \phi_i \phi_L \phi_k\cdots,\cdots)   \, A_R(\phi_R  \phi_j, \cdots) \, ,
\end{equation}
which again factorizes as $A(\cdots \phi_i \phi_j \phi_k\cdots, \cdots)$ should.

\medskip

\noindent {\it Scalars on Opposite Sides.}
The final case arises when the two biadjoint scalars are on opposite sides of the factorization channel.
Since $\phi_i$ and $\phi_k$ are by assumption within the same dual color trace, this means that a biadjoint scalar must be exchanged across the factorization channel
\begin{equation}
A(\cdots\phi_i \phi_k\cdots, g_j, \cdots) 
\sim A_L(\cdots\phi_i \phi_L\cdots, g_j,\cdots) \,   A_R(\phi_R \phi_k,\cdots) \, .
\end{equation}
${\cal T}_{ijk}$ only extracts the dependence on the momentum difference between $p_i$ and $p_k$.  However, only the total momenta through the factorization channel, $p_L$, so $p_k$ can only enter through the combination of momenta $p_L$.  This implies that on this factorization channel,
\eq{
	\partial_{p_i e_j } - \partial_{p_k e_j}  \rightarrow \partial_{p_i e_j } - \partial_{p_L e_j} \,  ,
}
which in terms of transmutation operators implies
\eq{
	{\cal T}_{ijk}\rightarrow {\cal T}_{ijL} \, .
}
Applying this to the factorization equation we obtain
\begin{equation}
{\cal T}_{ijk}\, A(\cdots\phi_i \phi_k\cdots, g_j,\cdots) \sim A_L(\cdots\phi_i \phi_j \phi_L\cdots,\cdots)  \, A_R(\phi_R \phi_k,\cdots) \, ,
\end{equation}
which factorizes in the same way as $A(\cdots \phi_i \phi_j \phi_k\cdots, \cdots)$.

\subsubsection*{Longitudinal Operator}

As shown in \Eq{eq:long_claim}, the longitudinal operator transmutes all the gluons in a single trace amplitude into pions, leaving spectator biadjoint scalars untouched.  To prove that the resulting amplitude is physical, it will be convenient to split  ${\cal L}$ according to its contributions coming from each side of the factorization channel,
\eq{
	{\cal L} = {\cal L}_L \cdot {\cal L}_R \qquad \textrm{where} \qquad {\cal L}_L = \prod_{i\in L} {\cal L}_i \qquad \textrm{and} \qquad  {\cal L}_R = \prod_{i\in R} {\cal L}_i \, .
	\label{eq:L_fact}
}
Note that the ${\cal L}_i$ factors still depend on momenta on both sides of the factorization channel, given the definition in \Eq{eq:long_def}.  As before, there are several possible configurations of the biadjoint scalars in the amplitude relative to the factorization channel, which we now consider in turn.

\medskip

\noindent {\it Scalars on Both Sides.}
By assumption, the starting amplitude is single trace.  Since there are biadjoint scalars on either side of the factorization channel, the exchanged particle must also be a biadjoint scalar, so
\eq{
	A(\cdots\phi_i \phi_l\cdots, g_j,  g_k,\cdots)\sim A_L(\cdots\phi_i \phi_L, g_j, \cdots)  \, A_R(\phi_R \phi_l \cdots, g_k,\cdots) \, ,
}
where $g_j$ and $g_k$ are by definition gluons on to the left and right of the factorization channel. 
Next, we consider a longitudinal operator ${\cal L}_j$ acting on gluon $g_j$.
Since the intermediate particle is a scalar, the polarization vector $e_j$ can only couple to $p_k$ for $k\in R$ through the combination given by the total momentum flowing through the channel, $p_I =\sum_{k \in R} p_k$. On the factorization channel, ${\cal L}_j$ is effectively equivalent to 
\eq{
	{\cal L}_j &= 
	\sum_{l \in L} p_l p_j \partial_{p_l e_j} + \sum_{k \in R} p_k p_j \partial_{p_k e_j}
	\rightarrow \sum_{l \in L} p_l p_j \partial_{p_l e_j} + p_I p_j \partial_{p_I e_j} = {\cal L}'_j \, .
	\label{eq:L_fact_easy}
}
where we effectively replace $\partial_{p_k e_j}$ with $\partial_{p_I e_j}$.
We recognize ${\cal L}'_j$ as a longitudinal operator acting purely on states in the left lower-point amplitude.  
Applying the same logic to the right, we obtain
\begin{equation}
\begin{split}
	{\cal L} \cdot A(\cdots\phi_i \phi_l\cdots, g_j, g_k,\cdots) &\sim {\cal L}_L \cdot A_L(\cdots\phi_i \phi_L, g_j, \cdots) \,   {\cal L}_R \cdot A_R(\phi_R \phi_l \cdots, g_k,\cdots) \\
	&\sim A_L(\cdots\phi_i \phi_L, \pi_j,\cdots)  \, A_R(\phi_R \phi_l \cdots,\pi_k,\cdots) \, ,
\end{split}
\end{equation}
which is the proper factorization equation for the transmuted amplitude, $A(\cdots\phi_i \phi_l\cdots, \pi_j,  \pi_k,\cdots)$.

\medskip

\noindent {\it Scalars on Same Side.}
When all of the biadjoint scalars are on the same side of factorization channel, the original amplitude factorizes with an intermediate gluon,
\eq{
	A(\cdots\phi_i \phi_l\cdots, g_j, g_k,\cdots) \sim A_L(\cdots\phi_i \phi_l\cdots ,  g_j, g_L,\cdots)  \, A_R(g_R, g_k,\cdots) \, .
	\label{eq:pion_case2}
}
Similar to the previous case, on the factorization channel, ${\cal L}_j$ is effectively equivalent to 
\begin{align}
{\cal L}_j
&\rightarrow {\cal L}'_j+ \sum_{q \in R} p_q p_j \partial_{p_q e_R} \cdot {\cal T}_{L j}  \, ,
\label{eq:Lleft_expand}
\end{align}
where ${\cal T}_{L j}=\partial_{e_L e_j}$ is the trace operator acting on $g_{j}$ and $g_L$. 
The additional term compared with Eq.~\eqref{eq:L_fact_easy} accounts for the fact that the intermediate particle is now a gluon, so the polarization vector $e_j$ can couple to $p_k$ through $p_k e_R$ via the completeness relation.
Likewise we have
\begin{align}
{\cal L}_k
&\rightarrow {\cal L}'_k+ \sum_{l \in L} p_l p_k \partial_{p_l e_L} \cdot {\cal T}_{R k} \,,
\end{align}
for $k\in R$.
Plugging the above into ${\cal L}_R$, this becomes the logitudinal operators for the right sub-amplitude except for the two gluons $g_{k,R}$, whose transmutation is given by the extra term in the above equation.
The trace operator contained there turns gluons into biadjoint scalars, which is exactly what we need to form pion amplitude on the right. Effectively,
\begin{equation}
{\cal L}_R \rightarrow
\sum_{k \in R, l \in L} p_k p_l \partial_{p_l e_L} \cdot ({\cal L}'_{R} \cdot {\cal T}_{R k}) = {\cal L}'_{I}\cdot ({\cal L}'_{R} \cdot {\cal T}_{R k}) \, ,
\end{equation}
where $ {\cal L}'_{I}$ is the longitudinal operator for the intermediate gluon on the left and $({\cal L}'_{R} \cdot {\cal T}_{R k})$ is the operator taking YM sub-amplitude into NLSM one.
To get the second equality, we use the fact that $({\cal L}'_{R} \cdot {\cal T}_{R k})$ yields the same pion amplitude for any $k$, so the summation on $k$ can be carried out using $\sum_{k \in R} p_k = p_I$, which then gives ${\cal L}'_{I}$. 
We find the ${\cal L}_R$ on the factorization channel simplifies into two parts: transforming the YM sub-amplitude on the right into a NLSM one as well as taking the internal gluon on the left into a pion.
Applying this to the full amplitude yields
\begin{equation}
\begin{split}
{\cal L} \cdot A(\cdots\phi_i \phi_l\cdots, g_j, g_k,\cdots) 
&\sim {\cal L}_L \cdot {\cal L}_R \cdot  A_L(\cdots\phi_i \phi_l\cdots ,  g_j, g_L,\cdots) \,    A_R(g_R, g_k,\cdots)  \\
&\sim {\cal L}_L \cdot {\cal L}'_I\cdot  A_L(\cdots\phi_i \phi_l\cdots ,  g_j, g_L,\cdots) \, A_R(\pi_R, \pi_k\cdots)  \\
&\sim  A_L(\cdots\phi_i \phi_l\cdots, \pi_j,\pi_L ,\cdots ) \, A_R(\pi_R , \pi_k\cdots)  \, ,
\end{split}
\end{equation}
which factorizes correctly with an internal pion. Going from second to third line, the second term in Eq.~\eqref{eq:Lleft_expand} is irrelevant because $g_L$ has been transmuted. The rest of operators in ${\cal L}_L {\cal L}'_I$ transmute gluons in $A_L$ into pions, yielding the desired factorization.

\subsection{Double Copy Construction}

Our induction proof establishes the unifying relations which are summarized in \Eq{eq:YM_sum} and depicted in the lower triangle of Fig.~\ref{fig:unified_web}.  To derive additional unifying relations, we simply employ the double copy construction in the spirit of \cite{Cachazo:2014xea}. Crucially, since transmutation is independent of the particular representation chosen for the amplitude, this procedure commutes with the KLT relations.  In particular, recall that the KLT construction dictates that
\begin{align}
	A_{\rm BI} = A_{\rm YM} \otimes \tilde{A}_{\rm NLSM} \qquad \textrm{and} \qquad
	A_{\rm G} = A_{\rm YM} \otimes \tilde{A}_{\rm YM} \, .
\end{align}
By taking the KLT product of the amplitudes in \Eq{eq:YM_sum} with those of YM theory and the NLSM, we obtain the unifying relations for extended gravity and BI shown in \Eq{eq:GR_sum} and \Eq{eq:BI_sum}, respectively, and represented by the other triangles in Fig.~\ref{fig:unified_web}.  This completes our proof.

%
%

\section{Infrared Structure}
\label{sec:soft}

Since transmutation is a functional relationship between scattering amplitudes, it applies in all kinematic regimes.  
Nevertheless, it will be enlightening to study this operation specifically in the soft limit, where certain universal structures arise.  In particular, the gluon and graviton scattering amplitudes famously obey soft theorems  taking the schematic form
\begin{equation}
A^{(n)} \xrightarrow{p_i \rightarrow 0} {\cal S}^{(i)} \cdot A^{(n-1)} \, ,
\label{eq:soft_original}
\end{equation}
in the limit where particle $i$ is soft.  Here we have defined the soft operator, ${\cal S}^{(i)}$, which acts on the lower-point amplitude $A^{(n-1)}$ to yield $A^{(n)}$ in the soft limit.  At leading order in the soft limit, ${\cal S}^{(i)}$ is a simple multiplicative factor, but at higher order it acts as a differential operator on the hard external kinematic data.  

Next, let us study the image of these soft theorems for gluons and gravitons under transmutation.  Consider a generic product of transmutation operators ${\cal T}^{(n)}$ acting on  $A^{(n)}$.  For later convenience, we factorize ${\cal T}^{(n)}$ into 
\eq{
{\cal T}^{(n)}={\cal T}^{(i)}\cdot  {\cal T}^{(n-1)} \, ,\label{eq:Tn}
}
where ${\cal T}^{(i)}$ transmutes the soft particle and ${\cal T}^{(n-1)}$ transmutes the remaining hard particles.  Since ${\cal T}^{(n-1)}$ and $A^{(n-1)}$ are by definition independent of the soft particle, they do not contain $e_i$ and so $[\, {\cal T}^{(i)} \, ,\, {\cal T}^{(n-1)} \, ]=0$ and ${\cal T}^{(i)} \cdot A^{(n-1)}=0$.  Using this observation, we apply the transmutation operator in \Eq{eq:Tn} to the right-hand side of Eq.~\eqref{eq:soft_original}, yielding
\begin{equation}
\begin{split}
{\cal T}^{(n)} \cdot A^{(n)}  \xrightarrow{p_i \rightarrow 0} {\cal T}^{(i)} \cdot {\cal T}^{(n-1)} \cdot {\cal S}^{(i)} \cdot A^{(n-1)}
= [ \, {\cal T}^{(i)} \cdot {\cal T}^{(n-1)}\, ,\, {\cal S}^{(i)} \, ]\cdot A^{(n-1)} \, .
\end{split}
\end{equation} 
The hard trace and insertion operators commute with the soft operator, so $[\, {\cal T}^{(n-1)}\, , \, {\cal S}^{(i)} \, ]=0$ and  the above equation further simplifies to
\begin{equation}
\begin{split}
{\cal T}^{(n)} \cdot A^{(n)}  \xrightarrow{p_i \rightarrow 0} [ \, {\cal T}^{(i)} \, ,\, {\cal S}^{(i)} \, ] \cdot {\cal T}^{(n-1)}\cdot A^{(n-1)} \, .
\end{split}
\label{eq:soft_transmutation}
\end{equation}
Since ${\cal T}^{(n-1)}\cdot A^{(n-1)}$ is by definition the transmuted lower-point amplitude, this implies that the transmuted soft operator is simply $ [ \, {\cal T}^{(i)} \, ,\, {\cal S}^{(i)} \, ]$.  In what follows let us discuss the transmuted cousins of the leading and subleading soft theorems.


\subsection{Soft Theorems at Leading Order}

Our jumping-off point is the  leading order soft graviton factor of Weinberg \cite{Weinberg:1965nx},
\eq{
	{\cal S}_{\rm G}^{(i)}= \sum_{ j \neq i}   \frac{ p_i e_j  \, p_i \tilde e_j }{ p_{i} p_j } \,  ,
}
where we have maintained a factorized form for the polarizations to emphasize our focus on extended gravity.  We obtain the leading soft gluon factor by taking the commutator of a transmutation operator and the leading soft graviton factor.  

Applying the trace operator, we find a vanishing commutator,
\eq{
	{\cal S}_{{\rm EM}}^{(ij)}  &= [\, {\cal T}_{ij} \, ,\, {\cal S}_{{\rm G}}^{(i)} \,] = 0 \, ,
}
which correctly establishes the vanishing of soft photon amplitudes in EM theory at leading order.
For the insertion operator, this reproduces the known leading soft gluon factor,
\eq{
	{\cal S}_{{\rm YM}}^{(ijk)}  &= [ \, {\cal T}_{ijk} \, ,\, {\cal S}_{{\rm G}}^{(j)} \, ] = \frac{ p_i e_j}{ p_i p_j } -\frac{ p_k e_j}{ p_k p_j } \, ,
}
corresponding to a soft particle $j$ adjacent to particles $i$ and $k$. The differential operators here act implicitly on the tilded polarizations. Note that the appearance of color ordering is given by the definition of the insertion operator.

Applying the trace operator to the soft gluon factor, we obtain
\eq{
	{\cal S}_{{\rm YMS}}^{(ijk|jl)}  &= [ \, {\cal T}_{jl} \, ,\, {\cal S}_{{\rm YM}}^{(ijk)} \, ] = 0 \, ,
}
where the vertical bar in the superscript separates the original and dual color orderings, so particle $j$ in the same dual color trace as particle $l$.  As expected, we find that soft scalar amplitudes of YMS also vanish at leading order.  Instead, if we apply the insertion operator again, we obtain the leading soft biadjoint scalar soft factor,
\eq{
	S_{{\rm BS}}^{(ijk|ljm)}  &= [ \, {\cal T}_{ljm} \, ,\,  	{\cal S}_{{\rm YM}}^{(ijk)} \, ]  = \frac{\Delta_{lim}}{ p_i p_j } -\frac{\Delta_{lkm}}{ p_k p_j } \, ,
}
where $\Delta_{ijk}=\delta_{ij} - \delta_{jk}$ and the differential operators act implicitly on the nontilded polarizations.  Here the soft factor is non-zero only when $i$ or $k$ are still adjacent to $j$ in the dual  color ordering.  Just like the leading soft factors for gluons and gravitons, the leading soft factor for biadjoint scalars can also be trivially derived from Feynman diagrams.

Last of all, let us consider the action of the longitudinal operator on the leading soft theorems.  As shown in \Eq{eq:Wi_to_Li}, the longitudinal operator acting on a particle can be written as ${\cal L}_i = {\cal W}_i + \Delta_i$, where ${\cal W}_i$ is the Ward identity operator and thus annihilates a physical amplitude.   Since the soft gluon and graviton factors are independent of $e_i e_j$ factors, they are annihilated by ${\cal L}_i$, so the transmuted soft factors are all vanishing.  In particular,
\eq{
	S_{\rm BI}^{(i)} = S_{\rm NLSM}^{(ijk)} =0 \, ,
}
which is expected since BI theory is a manifestly derivatively coupled theory and the NLSM exhibits the well-known Adler zero \cite{Adler:1964um}. This reconfirms the observation of \cite{Cheung:2016prv} that the ${\cal O}(p)$ vanishing of pion amplitudes follows directly from leading soft gluon theorem.  Note that the Adler zero derived above also applies to a general amplitude of pions coupled to biadjoint scalars, which is necessary for the proof in \Sec{sec:proof}.  Furthermore, a trivial corollary is that
\eq{
	S_{\rm SG}^{(i)} = S_{\rm DBI}^{(ij)} =0 \, ,
}
since the vanishing soft limits of BI and NLSM are inherited by their descendants, DBI and SG.


\subsection{Soft Theorems at Subleading Order}

Next, let us transmute  the soft gluon theorem at subleading order to learn its implications for the descendants of YM theory, which are the BS theory, YMS theory, and the NLSM.  
The soft gluon theorem, including leading and subleading contributions~\cite{Casali:2014xpa,Schwab:2014xua,Bern:2014vva,Broedel:2014fsa}, is
\begin{equation}
{\cal S}_{\rm YM}^{(ijk)} = \frac{p_i e_j -p_j J_i e_j}{p_i p_j}
-(i \leftrightarrow k) \, ,
\end{equation}
where  $p_j J_i  e_j =  p_j^\mu J_{i\mu \nu} e_j^\nu$ and we have defined
\eq{
J_{i\mu \nu} =  p_{i[\mu} \frac{\partial}{\partial{p_i^{\nu]}}}+ e_{i[\mu} \frac{\partial}{\partial{e_i^{\nu]}}} \, ,
}
which is the angular momentum operator of particle $i$, split between orbital and spin components.

At subleading order, the commutator of the trace operator and the soft gluon operator is
\eq{
{\cal S}^{(ijk|jl)}_{\rm YMS}= [ \, {\cal T}_{jl} \, ,\,  {\cal S}_{\rm YM}^{(ijk)} \, ]  =- \sum_{m\neq i,k}\delta_{lm}{\cal T}_{ilk}+\left \lbrace \frac{{\cal N}_{ijkl}}{p_i  p_j}
- \left( i \leftrightarrow k\right)	\right \rbrace,
\label{eq:soft_YMS}
}
where ${\cal N}_{ijkl}=\sum_{m\neq i} \left( \delta_{il} p_j e_m- \delta_{lm} p_j e_i \right) {\cal T}_{im}+
 \sum_{m\neq i,k} \delta_{il} p_j p_m  {\cal T}_{mik} $.
This is a new subleading soft theorem for YMS theory.
Instead taking the commutator of the insertion  operator, we obtain
\eq{
{\cal S}_{\rm BS}^{(ijk|ljm)} = [\, {\cal T}_{ljm} \, , \, {\cal S}_{\rm YM}^{(ijk)} \, ] 
=  \left\{  \frac{\Delta_{lim} \left(1+p_j  \frac{\partial}{  \partial {p_i} }
	\right)}{p_i p_j }
-\frac{p_j  e_i}{p_i  p_j} {\cal T}_{lim}
-\left( i \leftrightarrow k\right)  \right\}  - {\cal T}_{ilkm} \, ,
\label{eq:soft_BS}
}
which is a new soft theorem for a biadjoint scalar at subleading order.  Note that the above expression is valid even for biadjoint scalars coupled to gluons.

Finally, let us derive the subleading soft theorem for the NLSM.   Here it will be convenient to use the longitudinal operator in the form  ${\cal L}_{ij}=-p_i p_j {\cal T}_{ij}$.
 While \Eq{eq:soft_transmutation} requires $[ \, {\cal T}^{(n-1)} \, ,\, S^{(j)} \, ]=0$, this is not strictly speaking true when the transmutation involves longitudinal operators and the soft gluon operator at subleading order. 
However, the nonzero contribution to the commutator comes from terms of the form $[ \, {\cal T}^{(n-1)} \, , \, \partial_{p_l p_m} \, ]$, which can only arise from the orbital angular operator in the subleading contributions to the soft operator. 
From the form of the orbital angular momentum operator this implies that the soft polarization $e_j$ must be dotted into a momentum vector, and so this term is automatically annihilated by ${\cal L}_{ij}$.    Hence the commutator $[\, {\cal T}^{(n-1)} \, ,\, S^{(j)} \, ]$ effectively vanishes, so we can employ the results obtained for the trace and insertion operators.  In particular, exploiting that ${\cal L}_{ij}$ is simply a linear combination of ${\cal T}_{ij}$ operators, then from which we can read off the subleading soft theorem of the NLSM from that of YMS theory in \Eq{eq:soft_YMS}, yielding
\eq{
{\cal S}^{(ijk)}_{\rm NLSM} = \sum_{l\neq i,k} p_j p_l {\cal T}_{ilk} \, .
}
The presence of the insertion operator implies that the subleading soft theorem for pions involves a lower-point amplitude of a biadjoint scalar, in agreement with  the results of \cite{Cachazo:2016njl}.  

Naively, it should be a straightforward exercise to extend these results to the subleading soft graviton theorem~\cite{Cachazo:2014fwa,Schwab:2014xua,Afkhami-Jeddi:2014fia,Bern:2014vva,Broedel:2014fsa,Zlotnikov:2014sva,Kalousios:2014uva}.   This is of particular interest given the central role of the soft graviton as the stress tensor for a CFT whose correlators are flat space scattering amplitudes \cite{Kapec:2016jld,Cheung:2016iub}.  Unfortunately, there is an obstruction to this path since transmutation requires a extended gravity amplitude which incorporates the graviton, dilaton, and two-form.  To the best of our knowledge, no such factorized subleading soft theorem exists for a general representation of the amplitude~\cite{DiVecchia:2015oba,DiVecchia:2016amo}.  One exception to this comes from the SG theory, where transmutation acts symmetrically on the two polarizations associated with the graviton.  There the leading and subleading soft graviton operators commute with ${\cal L}_{ij}$, verifying the subleading vanishing ${\cal O}(p^2)$ soft behavior observed in \cite{Cheung:2014dqa}.  Furthermore, as shown in \cite{Cachazo:2016njl}, the subsubleading soft theorem can be extracted from the KLT relations.

\section{Outlook and Future Directions}
\label{sec:outlook}

We have derived a set of simple differential operators which transmute physical scattering amplitudes into new ones.  Applied repeatedly, transmutation then spawns the family tree of theories depicted in \Fig{fig:unified_web}.  Because all of these theories are descendants of extended gravity, they are bred with innate structures like the KLT and BCJ relations, the CHY formulation, and soft theorems. Our results offer several avenues for future work, which we now discuss.

First and foremost is the question of the {\it physical} meaning of transmutation.  While the trace operator is a trivial implementation of dimensional reduction, the insertion and longitudinal operators are somewhat murkier in interpretation.  In principle, one should be able to derive transmutation purely at the level of the Lagrangian.  If such a prescription exists, it will likely shed light on the symmetry and algebra for color-kinematic duality observed in \cite{Cheung:2016prv}.

Clarifying the nature of transmutation might give us clue about how to include fermions. The discussion in this paper crucially relies on amplitudes valid in arbitrary spacetime dimension, so naturally only bosons are considered. 
On the other hand, it is well-known that supersymmetric Ward identities also relate states of different spins, including fermions, albeit they only exists in certain dimensions. It would be of interest to consider the combination of transmutation and supersymmetric Ward identities. For instance, the commutators of the operators between them may lead to a way of applying transmutation on fermions. 

A second but related issue concerns what it even means to define a physical theory.  Historically, theories are defined by considerations of symmetry. YM theory and gravity are dictated by gauge invariance and general covariance \cite{Arkani-Hamed:2016rak}, while effective theories are controlled by nonlinearly realized symmetries.  From the perspective of the S-matrix, these same constraints enter through Lorentz invariance, locality, and infrared properties \cite{Cheung:2016drk,Rodina:2016jyz}.  We have found here that transmutation maps these otherwise unrelated concepts onto each other.  Nevertheless, not every transmuted theory is obviously controlled by a physical principle.  In particular, it would be interesting to see what dictates the structure of the BS theory in the spirit of \cite{Arkani-Hamed:2016rak,Rodina:2016jyz,Rodina:2016mbk}.

Last but not least is the question of generalizing our results to loop-level amplitudes.  Obviously, transmutation applies to integrands on the unitarity cuts where the integrands reduce to a product of tree-level amplitudes. Whether some vestige of transmutation persists for the integrand in generic loop momenta, or even further at the level of full amplitudes remains to be seen but clearly deserves further study.




\section*{Acknowledgment}
We would like to thank Henrik Johansson and Ellis Yuan for discussions which inspired this work, as well as Bo Feng, Song He, Yu-tin Huang, Jan Plefka,  Oliver Schlotterer, Stephen Stieberger, and Wadim Wormsbecher for helpful discussions. CHS and CW are grateful to the KITP, Santa Barbara, for hospitality during the final stages of this work. CC is supported by a Sloan Research Fellowship and CC, CHS, and CW are supported in part by a DOE Early Career Award under Grant No. DE-SC0010255 and by the NSF under Grant No. NSF PHY-1125915.

\bibliographystyle{utphys-modified}
\bibliography{unifyref}{}

\end{document}